# High-Efficiency Octave Bandwidth Rectifier for Electromagnetic Energy Harvesting

Haoming He, *Graduate member, IEEE*, Yilin Zhou, Zhongqi He, Yuhao Feng, and Changjun Liu, *Senior Member*, *IEEE*

*Abstract*—This letter presents the design and implementation of a compact high-efficiency octave microwave rectifier. A key highlight is the novel segmented impedance matching method, a unique approach that expands the rectifier bandwidth. The diode reactance is initially regulated by a series short-ended microstrip line. Impedance-compensated structures, characterized by varying admittance properties across an extensive frequency range, partition the operating frequency band into two segments based on the input impedance, thereby minimizing impedance variation. Ultimately, the input impedance is matched by a novel triple-band matching network. An octave rectifier was fabricated and measured. Results demonstrate that the rectifier achieves over 50% efficiency over 1.3-2.55 GHz (fractional bandwidth 64.9%) at 0 dBm RF input power. Even with a decrease in input power to −10 dBm, the rectifier maintains over 30% efficiency.

## I. INTRODUCTION

With the continuous advancement of technology and the growing need for environmental sustainability, energy harvesting technologies are becoming a key means of solving energy supply problems. Energy harvesting, as one of the new technologies, has a very promising application prospects [1][2][3]. Electromagnetic energy harvesting enables the construction of self-powered or near-self-powered systems, which are critical for applications such as Internet of Things devices [4][5], wireless sensor networks [6][7][8], and communication nodes in remote areas. Broadband high-efficiency rectifiers that can be applied at low power are the focus of research for better harvesting of electromagnetic energy in space. However, simultaneously realizing both lower input power and wider operating bandwidth has challenged the design of rectifiers.

Several proposals have been suggested to enhance the operational efficiency of the rectifiers. Non-uniform transmission lines are used for impedance matching of wideband rectifiers to achieve high rectification efficiency [9][10]. Non-uniform transmission lines are large and not conducive to device integration. On the other hand, the rectifiers using harmonic controlling circuits have been

This work was supported by NFSC under Grant U22A2015 and 62071316, Sichuan Science and Technology Program under Grant 2024YFHZ0282. (Corresponding author: Changjun Liu).
H. He, Y. Zhou, Z. He, and C. Liu are with the School of Electronics and Information Engineering, Sichuan University, Chengdu, 610064, and also with the Yibin Industrial Technology Research Institute of Sichuan University, Yibin 644000, China (email: cjliu@ieee.org).
Y. Feng is with the School of Electronic Science and Engineering, University of Electronic Science and Technology of China, Chengdu, China

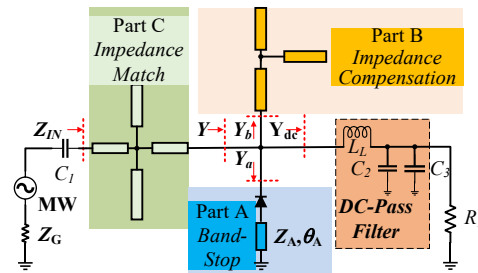

Fig. 1 Schematics of octave rectifier.

successfully enhanced, expanding the bandwidth without increasing the size [11][12]. In terms of novel impedance compression techniques, the primary method to achieve small size and high efficiency is to compress the variation range of the rectifier input impedance [13][14][15][16][17]. Using voltage-doubling circuits combined with broadband matching networks, such as simplified real-frequency techniques [18][19][20] and two-branch impedance matching circuits [21], can help make rectifiers compact and efficient.

This letter introduces an innovative octave rectifier with high efficiency for energy harvesting applications. The design includes impedance adjustment, compensation, and matching, which divides the rectifier's working frequency band into two segments: real impedance matching and complex impedance matching, improving both efficiency and operating bandwidth. We analyze the matching network and imaginary part compensation, then fabricate and measure an octave rectifier with an operating bandwidth of 1.3-2.55 GHz.

## II. PRINCIPLE AND DESIGN

Fig. 1 shows the operating principle diagram of the octave rectifier. It has three main parts: Part A adjusts the diode's admittance to $Y_a$, Part B supplies admittance $Y_b$ to compensate $Y_a$ to $Y$, and Part C matches the admittance $Y$ to the source. $Y_{dc}=0$, and $Y=Y_a+Y_b$.

In the design, to enhance the rectifier's bandwidth, the input admittance of the diode is optimized using two segments in its operating band. Define the starting and stopping frequencies of the operating band and the frequencies of the segments as fixed $f_1$, $f_3$, $f_2$, respectively, and $f_1 < f_2 < f_3$. It is required that the admittance of part A and B in parallel is complex from $f_1$ to $f_2$ and real from $f_2$ to $f_3$, which meets the design requirement of segment matching. Part C comprises a cross structure intended for impedance matching, meticulously designed to align the input impedance with the 50 Ω system source. The operating bandwidth of the rectifier is effectively broadened by matching the input impedance of the rectifier in two segments.

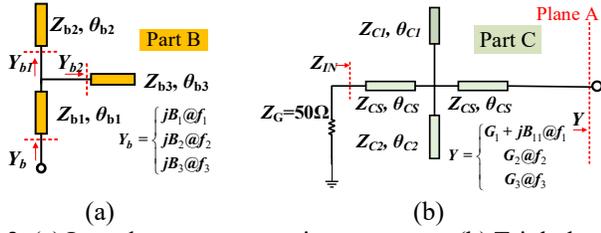

Fig. 2. (a) Impedance compensation structures. (b) Triple-band matching network.

The operating principles of Part B and C are analyzed initially. The microstrip line characteristic impedances and electrical lengths are both defined at $f_1$.

### A. Impedance Compensation Structures

The input impedance of diodes in rectifiers exhibits nonlinear variation and requires compensation for its imaginary component to effectively align with the matching network. The topology of impedance compensation structures is shown in Fig. 2 (a). $Y_{b1}$ and $Y_{b2}$ are the input admittance of microstrip lines. We denote $k_2 = f_2/f_1$ and $k_3 = f_3/f_1$.

To simplify the procedure, first, it is satisfied with $Y_{b2} = 0@f_2$, which could be achieved by implementing a quarter-wave open-ended transmission line. Subsequently, upon selecting and fixing the characteristic impedance of $Z_{B1}$, the electrical length of $\theta_{b1}$ is expressed as:

$$\theta_{b1} = \frac{1}{k_2}\tan^{-1}\left(\frac{-1}{Z_{b1}B_2}\right) \quad (1)$$

When $Z_{b1}$ and $\theta_{b1}$ are calibrated, the admittance at $f_1$ and $f_3$ is

$$Y_{b1} + Y_{b2} = \left[\frac{-j(B_1 Z_{b1} + \tan\theta_{b1})}{Z_{b1} + Z_{b1}^2 \tan\theta_{b1}}\right]_{f=f_1}$$
$$= \left[\frac{-j(B_3 Z_{b1} + \tan(k_3\theta_{b1}))}{Z_{b1} + Z_{b1}^2 B_3 \tan(k_3\theta_{b1})}\right]_{f=f_3} \quad (2)$$

It also satisfies

$$Y_{b1} + Y_{b2} = \left[\frac{j\tan(\theta_{b2})}{Z_{b2}} + \frac{j\tan(\theta_{b3})}{Z_{b3}}\right]_{f=f_1}$$
$$= \left[\frac{j\tan(k_3\theta_{b2})}{Z_{b2}} + \frac{j\tan(k_3\theta_{b3})}{Z_{b3}}\right]_{f=f_3} \quad (3)$$

Solving (3) and (4) in association, the structural parameters of the entire Part B can be calculated.

### B. Impedance Match Structure

An appropriate impedance matching circuit is essential for broadening the bandwidth of the rectifier. A novel triple-band matching network is proposed. As depicted in Fig. 2 (b), the triple-frequency matching network comprises a cross-shaped microstrip network structure. When working with three frequencies, $f_1$, $f_2$, and $f_3$, it can realize impedance matching at $f_2$ and $f_3$ for pure resistance, as well as realize impedance matching at $f_1$ for complex impedance.

A quarter wavelength transmission line is often used as the impedance converter in impedance matching. The relationship between input impedance $Z_{in}$ and load impedance $Z_L$ meets

$$Z_{in} = \frac{Z_0^2}{Z_L} \quad (4)$$

where $Z_0$ is the transmission line characteristic impedance.

For frequencies $f_2$ and $f_3$, the cross circuit is equivalent to a quarter-wave transmission line, so both have the same ABCD matrix. ABCD matrix for part C is as follows:

$$\begin{bmatrix} A & B \\ C & D \end{bmatrix} =$$
$$\begin{bmatrix} \cos 2\theta_{CS} - \frac{1}{2}T_0 Z_{CS}\sin 2\theta_{CS} & jZ_{CS}\left(\sin 2\theta_{CS} - T_0 Z_{CS}\sin^2\theta_{CS}\right) \\ j\frac{1}{Z_{CS}}\left(\sin 2\theta_{CS} + T_0 Z_{CS}\cos^2\theta_{CS}\right) & \cos 2\theta_{CS} - \frac{1}{2}T_0 Z_{CS}\sin 2\theta_{CS} \end{bmatrix} \quad (5)$$

$$T_0 = \left(\frac{\tan\theta_{C1}}{Z_{C1}} + \frac{\tan\theta_{C2}}{Z_{C2}}\right) \quad (6)$$

Where $Z_{CS}$, $Z_{C1}$, $Z_{C2}$, $\theta_{CS}$, $\theta_{C1}$, $\theta_{C2}$ are the characteristic impedance and electrical length of different microstrip lines, as shown in Fig. 2 (b).

The ABCD matrix of a $\lambda_0/4$ transmission line is

$$\begin{bmatrix} A_0 & B_0 \\ C_0 & D_0 \end{bmatrix} = \begin{bmatrix} 0 & jZ_0 \\ j\frac{1}{Z_0} & 0 \end{bmatrix} \quad (7)$$

When $A=D=0$, we can obtain

$$\frac{1}{Z_{CS}}\left(\frac{\cos^2\theta_{CS} - \sin^2\theta_{CS}}{\sin\theta_{CS}\cos\theta_{CS}}\right) = \frac{\tan\theta_{C1}}{Z_{C1}} + \frac{\tan\theta_{C2}}{Z_{C2}} \quad (8)$$

We substitute B and C into the chain matrix and obtain

$$Z_0 = Z_S \tan\theta_S \quad (9)$$

Cross circuits have characteristic impedances $Z_{02}$ and $Z_{03}$ at $f_2$ and $f_3$ [22]

$$Z_{02} = Z_{CS}\tan k_2\theta_{CS} \quad (10)$$
$$Z_{03} = Z_{CS}\tan k_3\theta_{CS} \quad (11)$$

Because the admittances at $f_2$ and $f_3$ looking to the right in plane A are $G_2$ and $G_3$, respectively, we have

$$Z_{02}^2 = \frac{Z_{IN}}{G_2} \quad Z_{03}^2 = \frac{Z_{IN}}{G_3} \quad (12)$$

Solving (10)-(12), we can solve for $Z_{CS}$ and $\theta_{CS}$. At frequency $f_1$, the ABCD matrix looking to the right from plane A are

$$\begin{bmatrix} A_{f1} & B_{f1} \\ C_{f1} & D_{f1} \end{bmatrix} = \begin{bmatrix} A & B \\ C & D \end{bmatrix}\begin{bmatrix} 1 & 0 \\ G_{IN} & 1 \end{bmatrix} = \begin{bmatrix} A+BG_{IN} & B \\ C+DG_{IN} & D \end{bmatrix} \quad (13)$$

where $G_{IN}=1/Z_{IN}$. Therefore, at $f_1$, it obtains

$$Y_1 = \frac{C_{f1}}{A_{f1}} = \frac{C+DG_{IN}}{A+BG_{IN}} = G_1 + jB_{11} \quad (14)$$

Based on the derivation, the designed triple-band matching network can realize pure resistance impedance matching at $f_2$ and $f_3$ and complex impedance matching at $f_1$.

### C. Design of the Octave Rectifier

The design of an octave rectifier is underway after analyzing the matching network and impedance compensation structure. The design uses an SMS7630 diode with SPICE parameters shown in Table I. Firstly, Part A is a band-stop structure[23], which can reflect the high harmonics, and adjust the impedance of the diode to simplify subsequent design. The impedance $Z_a$ ($Z_a=1/Y_a$) curve of the diode after series connection with Part A at different frequencies is shown in Fig. 3 (a). The frequency intersecting with the real axis of the Smith chart is as $f_2$, $f_2=2.05$ GHz, and $f_1=1.3$ GHz, $f_3=2.6$ GHz. Then determining the admittances $Y_a$ corresponding to the $f_1$, $f_2$, and $f_3$, design and optimize the impedance compensation structure and matching network according to $Y_a$. The impedance variation curve $Z$ ($Z=1/Y$) of the rectifier in parallel with Part B is shown in Fig. 3 (b). It is seen that the input impedance of the rectifier is zero in the imaginary part from $f_2$ to $f_3$, whereas it is negative in the frequency range of $f_1$ to $f_2$. The real parts of the rectifier input impedance at $f_2$ and $f_3$ are around 120 Ω and 280 Ω, respectively, which are brought into (10)-(12) to obtain $Z_{CS}≈100$ Ω and $θ_{cs}≈λ_g/11$. After optimizing the design, the input impedance $Z_{IN}$ of the rectifier after cascading with matching network Part C is shown in Fig. 3 (c), and the rectifier is well matched with the signal source.

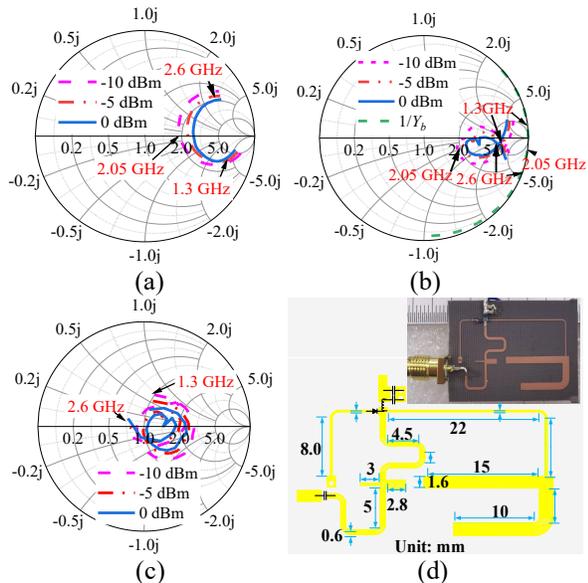

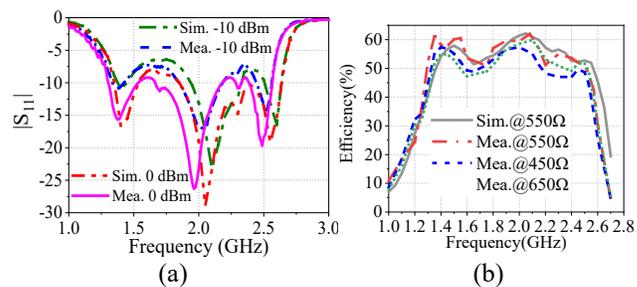

Fig. 4. (a) |$S_{11}$| with various frequencies. (b) PCE versus frequency with different dc-load at 0 dBm.

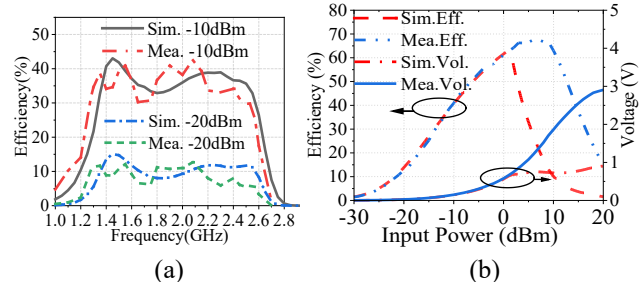

Fig. 3. (a) $Z_a$, (b) $Z$ ($Z=1/Y$), and (c) $Z_{IN}$ (d) The layout and photograph of the proposed octave rectifier.

Fig. 5. (a) PCE at −10 dBm and −20 dBm. (b) PCE versus input power at 2.08 GHz.

TABLE I
SPICE PARAMERERS OF A SMS7630 DIODE

| $B_v$ | $C_{j0}$ | $I_s$ | $R_s$ | $V_{bi}$ |
|---|---|---|---|---|
| 2.0 V | 0.14 pF | 5 μA | 20 Ω | 0.34 V |

TABLE II
COMPARIson WITH PREVIOUS RECTIFIERS

| Ref | Diode Models and Number | Bandwidth range for PCE >50% | | Power (dBm) | Size (mm$^2$) |
|---|---|---|---|---|---|
| | | Frequency range (GHz) | Relative bandwidth (%) | | |
| [12] | BAT15-03W,1 | 1.77-2.88 | 47.7 | 0 | 23×23 |
| [14] | HSMS286,1 | 2-3.3 | 49.1 | 4 | 31×18 |
| [15] | SMS7630, 2 | 1-1.7 | 51.9 | 0 | 35×25 |
| [16] | HSMS286, 1 | 1.5-2.85 | 49.0 | 5 | 37×10 |
| [19] | SMS7630,4 | 1.6-3.4 | 66.6 | 8 | 22×24 |
| This Work | SMS7630, 1 | 1.3-2.55 | 64.9 | 0 | 32×21 |

### III. MEASUREMENT AND ANALYSIS

To verify the feasibility of the proposed octave rectifier design theory, we designed and fabricated an octave rectifier with an operating bandwidth of 1.3 to 2.6 GHz and measured its rectification efficiency. F4B-2 is the substrate for the rectifier, with a relative dielectric constant of 2.65. The layout and photograph octave rectifier are illustrated in Fig. 3 (d). $C_1$ (39 pF) is a dc-block capacitor. The dc-pass filter comprises an inductor $L$ (22 nH) along with two capacitors, $C_2$ (13 pF) and $C_3$ (39 pF).

Fig. 4 (a) displays the simulated and measured reflection coefficient |$S_{11}$| with different frequencies at 0 dBm and 10 dBm. The |$S_{11}$| is less than -10 dB at 0 dBm, which is well matched. As depicted in Fig. 4 (b), the conversion efficiency is simulated and measured against the frequency for an input power of 0 dBm with different dc-loads. The rectifier achieves a PCE of more than 50% over 1.3 GHz-2.55 GHz with 550 Ω. It is reaching an octave level with a relative bandwidth of 64.9%. The measurement results at lower input power levels are shown in Fig. 5 (a). At −10 dBm, the rectifier has a PCE of more than 30% in the octave range of 1.25 GHz -2.5 GHz. Fig. 5 (b) depicts the rectifier operating optimally at 2.08 GHz, obtaining a maximum rectification efficiency of 67.5% at 6 dBm. The higher measured efficiency at the large-power state is mainly due to the higher reverse breakdown voltage of the actual diode compared to the simulation diode model.

Table II displays a comparative analysis of the rectifier's performance compared to previous studies. The findings reveal that the proposed rectifier achieves a fractional bandwidth (with PCE > 50%) of 64.9%, the highest among the compared studies at 0 dBm, and it utilizes only one diode.

### V. CONCLUSION

In this letter, a low-power and high-efficiency octave rectifier is proposed and designed for electromagnetic energy harvesting. The design is successfully broadened by partitioning the operating frequency band of the rectifier into two segments, resistance and impedance matching, in accordance with the impedance variation. Through theoretical analysis and practical measurement, the rectifier has a more than 50% rectification efficiency between 1.3 to 2.55 GHz. Even with a decrease in input power to -10 dBm, it maintains an efficiency level exceeding 30%. The rectifier design is applicable for electromagnetic energy harvesting and wireless power transfer across various applications.

## REFERENCES

[1] N. Shinohara, Trends in Wireless Power Transfer: WPT Technology for Energy Harvesting, Mllimeter-Wave/THz Rectennas, MIMO-WPT, and Advances in Near-Field WPT Applications," *IEEE Microw. Mag.*, vol. 22, no. 1, pp. 46-59, Jan. 2021.

[2] Y. Huang, N. Shinohara, and T. Mitani, "Impedance matching in wireless power transfer," *IEEE Trans. Microw. Theory Techn.*, vol. 65, no. 2, pp. 582–590, Feb. 2017.

[3] C. Liu, H. Lin, Z. He and Z. Chen, "Compact Patch Rectennas Without Impedance Matching Network for Wireless Power Transmission," *IEEE Trans. Microw. Theory Techn.*, vol. 70, no. 5, pp. 2882-2890, May 2022.

[4] H. Lin, X. Chen, Z. He, Y. Xiao, W. Che and C. Liu, "Wide Input Power Range X-Band Rectifier with Dynamic Capacitive Self-Compensation," *IEEE Microw. Wireless Compon.Lett*, vol. 31, no. 5, pp. 525-528, May 2021.

[5] P. Kamalinejad et al., "Wireless energy harvesting for the Internet of Things," *IEEE Commun. Mag.,* vol. 53, no. 6, pp. 102–108, Jun. 2015.

[6] P. Wu, S. -P. Gao, Y. -D. Chen, Z. H. Ren, P. Yu and Y. Guo, "Harmonic-Based Integrated Rectifier–Transmitter for Uncompromised Harvesting and Low-Power Uplink," *IEEE Trans. Microw. Theory Techn*, vol. 71, no. 2, pp. 870-880, Feb. 2023.

[7] S. Kim et al., "Ambient RF energy-harvesting technologies for self-sustainable standalone wireless sensor platforms," *Proc. IEEE*, vol. 102, no. 11, pp. 1649–1666, Nov. 2014.

[8] G. C. Martins and W. A. Serdijn, "An RF energy harvesting and power management unit operating over −24 to +15 dBm input range," *IEEE Trans. Circuits Syst. I, Reg. Papers*, vol. 68, no. 3, pp. 1342–1353, Mar. 2021.

[9] J. Kimionis, A. Collado, M. M. Tentzeris, and A. Georgiadis, "Octave and decade printed UWB rectifiers based on nonuniform transmission lines for energy harvesting," *IEEE Trans. Microw. Theory Techn.*, vol. 65, no. 11, pp. 4326–4334, Nov. 2017.

[10] P. Wu et al., "Compact high-efficiency wideband rectifier with multistage-transmission-line matching," *IEEE Trans. Circuits Syst. II, Exp. Letters*, vol. 66, no. 8, pp. 1316–1320, Aug. 2019.

[11] D.-A. Nguyen and C. Seo, "Design of high-efficiency wideband rectifier with harmonic control for wireless power transfer and energy harvesting," *IEEE Microw. Wireless Compon. Lett.*, vol. 32, no. 10, pp. 1231–1234, Oct. 2022.

[12] G. T. Bui, D. -A. Nguyen and C. Seo," A highly efficient design of wideband rectifier with harmonic suppression transferring for energy harvesting and wireless power transfer," *IEEE Microw. Wireless Compon.Lett.*, vol. 33, no. 7, pp. 1059-1062, July 2023.

[13] S. F. Bo, J. -H. Ou and X. Y. Zhang, "Ultrawideband Rectifier with Extended Dynamic-Power-Range Based on Wideband Impedance Compression Network," *IEEE Trans.Microw. Theory Techn*, vol. 70, no. 8, pp. 4026-4035, Aug. 2022.

[14] Z. He and C. Liu, "A compact high-efficiency wideband rectifier with a wide dynamic range of input power for energy harvesting," *IEEE Microw. Wireless Compon. Lett.*, vol. 30, no. 4, pp. 433–436, Apr. 2020.

[15] W. Liu, K. Huang, T. Wang, J. Hou and Z. Zhang, "Broadband High-Efficiency RF Rectifier with a Cross-Shaped Match Stub of Two One-Eighth-Wavelength Transmission Lines," *IEEE Microw. Wireless Compon. Lett.*, vol. 31, no. 10, pp. 1170-1173, Oct. 2021.

[16] H. He, H. Lin, P. Wu, Q. Li and C. Liu, "Compact High-Efficiency Wideband Rectifier Based on Coupled Transmission Line," *IEEE Trans. Circuits Syst. II, Exp. Letters,* vol. 69, no. 11, pp. 4404-4408, Nov. 2022.

[17] H. S. Park and S. K. Hong, "Wideband RF-to-dc rectifier with uncomplicated matching network," *IEEE Microw. Wireless Compon. Lett.*, vol. 30, no. 1, pp. 43–46, Jan. 2020.

[18] W. Liu, K. Huang, T. Wang, J. Hou and Z. Zhang, "A Compact High-Efficiency RF Rectifier with Widen Bandwidth," *IEEE Microw. Wireless Compon. Lett.*, vol. 32, no. 1, pp. 84-87, Jan. 2022.

[19] W. Liu, K. Huang, T. Wang, J. Hou and Z. Zhang, "A Compact Ultra-Broadband RF Rectifier Using Dickson Charge Pump," *IEEE Microw. Wireless Compon. Lett.*, vol. 32, no. 6, pp. 591-594, June 2022.

[20] Long H J, Cheng F, Yu S, et al. High-efficiency broadband rectifier with compact size for wireless power transfer. *Micro and Optic Techn Lett*, 64(11): 2007-2013, 2022.

[21] S. Zheng, W. Liu, and Y. Pan, "Design of an ultra-wideband high-efficiency rectifier for wireless power transmission and harvesting applications," *IEEE Trans. Ind. Informat.*, vol. 15, no. 6, pp. 3334–3342, Jun. 2019.

[22] Hu Z, Huang C, He S, et al. Tri-band matching technique based on characteristic impedance transformers for concurrent tri-band power amplifiers design. *2015 IEEE Region 10 Conference*, Macao. 2015.

[23] C. Liu, F. Tan, H. Zhang, and Q. He, "A novel single-diode microwave rectifier with a series band-stop structure," *IEEE Trans. Microw. Theory Techn.*, vol. 65, no. 2, pp. 600–606, Feb. 2017.